\documentclass[footinbib,prx,twocolumn,amsmath,amssymb,aps,superscriptaddress]{revtex4-1}
\usepackage{graphicx}
\usepackage{hyperref}

\hypersetup{
     colorlinks   = true,
     linkcolor    = blue,
     citecolor    = blue
}

\usepackage{amsthm}
\usepackage{physics}

\begin{document}

\title{Irreducible multi-partite correlations as an order parameter for k-local nontrivial states}
\author{Yahya Alavirad}
\thanks{The authors contributed equally to this work.}
\affiliation{Department of Physics, University of California at San Diego, La Jolla, CA 92093, USA}
\author{Ali Lavasani}
\thanks{The authors contributed equally to this work.}
\affiliation{Condensed Matter Theory Center, University of Maryland, College Park, Maryland 20742, USA}
\affiliation{Joint Quantum Institute, University of Maryland, College Park, Maryland 20742, USA}

\begin{abstract}
 Geometrically nontrivial quantum states can be defined as states that cannot be prepared by a constant depth geometrically local unitary circuit starting from a product state.  However, for topological phases, as well as a large class of quantum error correcting codes without an underlying geometric structure, the required circuit depth remains infinite even if we replace the condition of geometric locality with the weaker condition of k-locality. Motivated by this observation, we look for a non-geometric quantity that can capture k-local non-triviality of a given state, for example, we ask if it is possible to distinguish the ground state of the toric code from a trivial state without having access to the position label of the qubits. We observe that a fundamental property of k-local nontrivial states is the presence of irreducible many-partite correlations shared between an infinitely large number of randomly chosen parties, i.e. correlations that cannot be inferred by accessing only a finite number of parties. We introduce an order parameter designed to capture such correlations. We demonstrate the utility of our order parameter by applying it to a wide variety of examples: The toric code on a square lattice, random stabilizer states, quantum expander codes, and a particular holographic stabilizer state. We discuss general relations between this order parameter and the erasure thresholds of quantum error correcting codes as well as the classical bond percolation problem.

 \end{abstract}

\maketitle

\section{Introduction}

 Understanding and classifying quantum phases of matter is one of the major goals of theoretical condensed-matter physics. Non-trivial quantum phases are sometimes characterized using the ``quantum circuit definition": States belonging to a $d$ dimensional non-trivial phase cannot be obtained from the product state using a finite depth \textit{geometrically} local ($d$ dimensional) unitary quantum circuit \cite{chenwen,chenwenbook}.

 However, intriguingly, for (non-invertible) topologically ordered states the minimum circuit depth required remains infinite even if we replace the condition of \textit{geometric locality} and demand only the much weaker condition of \textit{k-locality} ~\cite{pcp,aharonov2018quantum}. That is, if we only consider unitary gates that act on at most $k$ qubits at a time, without any restriction on their relative placement (as opposed to unitary gates that act on a geometrically local region), we still cannot transform trivial states into topological states in a finite depth. This suggests that the classification of quantum phases of matter can be extended beyond the geometrically-local phases to the k-local setting. In addition to conventional topological states, examples of nontrivial k-local states \textit{without} any underlying geometrical structure are prevalent in the quantum information context, e.g.  quantum LDPC codes~\cite{gottesman2014faulttolerant,breuckmann2021ldpc} and random quantum codes~\cite{brown2013short,gullans2020quantum}.

 Given their relevance to both condensed matter and quantum information physics, we initiate the study of such ``k-local" nontrivial states. In particular, we ask: is it possible to construct an entirely \textit{non-geometric} order parameter for such k-local non-trivial states? For example, is it possible to distinguish the ground state of the toric code from a typical state in the trivial phase without having access to the underlying geometry, i.e. by removing the position label of all qubits? Or, is it possible to distinguish an encoded state of a quantum LDPC code from a typical trivial state without having access to the underlying interaction graph? We show that, at least in a large class of examples, this question can be answered in the positive. We argue that irreducible multi-partite correlations amongst random subsystems (at a suitably chosen sampling rate) can serve as an order parameter for such k-local nontrivial states. We will use this feature to define a non-geometric order parameter for k-local nontrivial states and numerically demonstrate that it can be used to detect a phase transition from a nontrivial phase to a trivial phase.

 Using a combination of numerical and analytical methods, we study several examples representing a wide class of k-local non-trivial states including geometric topological phases (toric code), random stabilizers states, quantum LPDC codes (quantum expander codes~\cite{leverrier2015quantum}), and Holographic error correcting codes (the holographic hexagon state~\cite{pastawski2015holographic}).

 In the case of random stabilizer states, we provide an analytical computation of our order parameter (complemented by numerics) and demonstrate that it obeys a set of scaling relations. We find that similar scaling relations are also applicable in all the rest of the examples. We extract a set of corresponding exponents from the scaling relation. An extensive theoretical study of the underlying theory explaining these exponents is left for future work.

 For geometric topological phases, we provide a way to understand our order parameter using its relation to the classical problem of bond-percolation. We present a detailed numerical study in the case of the toric code that confirms our percolation picture. We study our order parameter across a phases transition from the topological phase to the trivial phase and establish that the phase transition can be detected without having access to the underlying geometric structure. Additional examples, i.e. quantum expander codes, and holographic stabilizer states are also studied numerically, yielding similar results.

 The rest of this article is organized as follows. In Section \ref{sec2}, we define Irreducible multipartite correlations and discuss their basic properties. In Section \ref{opm}, we define our parameter and discuss the intuition behind it. In Section~\ref{secex}, we discuss numerical and analytical results for particular examples: random stabilizer states and the toric code as well a phase transition in random hybrid quantum circuits, that showcase the utility of our order parameter. Additional case studies of quantum expander codes and Holographic quantum error correcting states are presented in Appendices \ref{apx_expander_review}, \ref{holest} respectively. In \ref{secerase}, we discuss a relation between our order parameter and erasure threshold in error correcting quantum codes. We conclude with a brief discussion in Section \ref{secdics}.

\section{Irreducible multipartite correlations}\label{sec2}
Consider a density matrix $\rho$ shared between a set of $M$ non-overlapping parties $A_1,\cdots,A_M$. The marginal density matrix $\rho_j$ is defined as the reduced density matrix obtained by tracing out $A_j$, i.e. $\rho_j=\Tr_{A_j}\rho$.$A_1,\cdots,A_M$ have non-zero irreducible M-partite correlations if there is more information in $\rho$ than in the set of its marginals $\{\rho_j\}_{j=1}^M$.

Let $\tilde{\rho}$ denote the \textit{maximum entropy state} consistent with all marginals of $\rho$, meaning $\tr_{A_j}\tilde{\rho}=\rho_j$ for all $j$. As discussed in Refs.~\cite{wootters,kato,zhou,zhou2}, a quantitative measure of irreducible M-partite correlations can then be defined as:
\begin{align}\label{pM}
D_M(\rho,\mathcal{A})=S(\tilde{\rho})-S(\rho),
\end{align}
where the function $S(\rho)=-\Tr (\rho \log_2{\rho})$ is the von-Neumann entanglement entropy and $\mathcal{A}=\{A_1,\cdots,A_M\}$ denotes the partitioning. Intuitively, this object quantifies how many bits of information in $\rho$ cannot be read-off unless \textit{all} $M$ parties are accessed. Perhaps, the simplest example of a state with $D_M>0$ is provided by the GHZ state~\cite{wootters,Bravyi06}.

\subsection{Irreducible multipartite correlations in stabilizer states}\label{dstab}

Computing $D_M$ is not easy in general. However, for stabilizer states it can be efficiently computed. Let $A$ denote a set of $n_A$ qubits in a mixed stabilizer state specified by the stabilizer density matrix $\rho_A$ given as
\begin{align}
\rho_A=\frac{1}{2^{n_A}}\sum_{g\in G_A} g.
\end{align}
where $G_A$ is a the stabilizer group corresponding to $\rho_A$. The entanglement entropy of $\rho$ can be related to its stabilizer group as follows\cite{fattal2004entanglement},
\begin{align}\label{eq_stabee}
  S(\rho)=n_A-\dim G_A.
\end{align}
Let $\mathcal{A}$ denote a partitioning of $A$ into a set of $M$ non-overlapping parties (subsystems)  $A=\cup_{j=1}^M A_j$. $D_M$ can then be equivalently written as~\cite{Bravyi06,zhou},
\begin{align}\label{pk}
D_M(\rho,\mathcal{A})=\text{dim}~G_A-\text{dim}~G_{\mathcal{A};(M-1)},
\end{align}
where $G_{\mathcal{A};(M-1)}$ is defined as the stabilizer group generated by stabilizers in $G_A$ which are shared by at most $(M-1)$ parties in $\mathcal{A}$. The generator set for $G_{\mathcal{A};(M-1)}$ can be written as $\cup_{j=1}^M G_{A\setminus A_j}$.

In the case where the system studied is pure $S(\rho_A)=0$, it was shown in Ref.~\cite{Bravyi06} that $D_M(\rho,\mathcal{A})$ is equal to the $M$ partite GHZ extraction yield of the stabilizer state with respect to $\mathcal{A}$, that is, the maximum number of $M$ partite GHZ states that can be extracted by local unitaries (local to each party). We further remark that if the $M$-partite GHZ extraction yield of stabilizer states is non-zero for a macroscopic $M$, i.e. for $M$ growing with system size, it immediately follows that the state cannot be prepared with a constant depth k-local unitary circuit and is therefore k-local nontrivial.
\begin{figure}
  \includegraphics[width=\columnwidth]{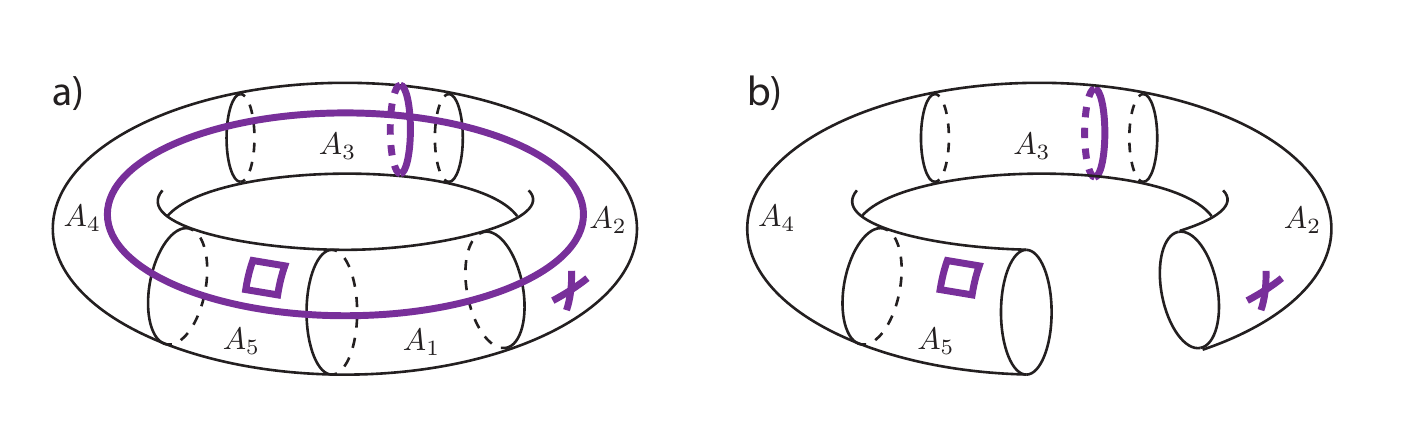}
  \caption{a) The toric code state on a torus. Thick lines represent some of its stabilizers. The system is partitioned into five disjoint parties $A=\cup_{j=1}^5 A_j$. b) The marginal $A\setminus A_1$. All stabilizers except the one that goes around the large handle of the torus can be read from the marginals, resulting in $D_5=1$.}
  \label{fig_torus}
\end{figure}

To develop intuition, let us consider the multipartite correlations shared between different regions (subsystems) of the ground state of the Kitaev's toric code on a torus (or any other $2D$ topological phase), with logical stabilizers going around both the non-contractible loops of the torus. As a simple example, divide the torus into a set of $M$ equal size ``cake-slices" corresponding to $A_1,\cdots,A_M$ (see Fig.\ref{fig_torus}). Apart from the logical qubit that goes around the torus, all of the information is encoded locally and therefore present in $M-1$ party density matrices (all the other stabilizer generators or Wilson loops can be taken to have support only in one region or in the boundary of two). However, the logical operator that goes around the cake  cannot be accessed without having \textit{all} $M$ of cake slices (cannot be contracted to any $M-1$ of the subsystems). We, therefore, have $D_M=1$.

  \subsection{Upper bound from conditional quantum mutual information}
  We can use the strong subadditivity of von-Neumann entropy to upper bound $D_M$. Consider dividing the $M$ parties in $A$ into any three disjoint sets $B$, $C$ and $D$ such that $BC$, $CD$ and $BD$ are all subsets of marginals of $A$ - e.g. $B=\cup_{i=1}^{M-2} A_i$, $C=A_{M-1}$ and $D=A_{M}$. Let $\tilde \rho$ be the maximum entropy state consistent with all marginals of $\rho$. Strong subadditivity for $\tilde \rho$ entails,
  \begin{align}\label{eq_ssa}
    S(\tilde \rho)\le S(\tr_{C}(\tilde \rho)) + S(\tr_{D}(\tilde \rho))- S(\tr_{CD}(\tilde \rho)).
  \end{align}
  Since $\tilde \rho$ has the same marginals as $\rho$, we can compute the entropies on the right hand side on $\rho$ instead of $\tilde \rho$.
  Therefore, we have the following upper bound on $D_M$
  \begin{align}\label{bound}
 D_M(\rho,\mathcal{A})&\le S(BD)+S(BC)-S(B)-S(BCD)\nonumber \\
                        &=I(C:D|B),
  \end{align}
  where $I(C:D|B)$ is the conditional quantum mutual information, and in an abuse of notation, we have used $S(X)$ to denote $S(\tr_{A \setminus X} \rho)$ for a subset $X$ of the qubits. Note that all entropies on the right hand side correspond to $\rho$ rather than $\tilde \rho$ and hence are easily computable.

 \section{Irreducible multipartite correlations between random subsystems in different quantum phases of matter}\label{opm}

  We now define our proposed quantity of interest. We start with the explicit definition and then explain the intuition behind it.

  Consider a quantum state of $N$ qubits. For a given $M$ and $q$ such that $0<Mq \le 1$, choose the subset $A$ and its $M$-partitioning as follows; Label every qubit with a random number $x_i$ chosen uniformly from $[0,1]$. Define $A_j$ as the set of qubits with labels between $(j-1)q$ and $j~q$ and let $A$ denotes union of $A_1$ through $A_M$, i.e.
\begin{align}
  A_j&=\{i:~(j-1)\,q < x_i <j\,q\} \quad\text{for } j=1,\cdots,M~, \\
  A&=\cup_{j=1}^M A_j,\label{eq_A}\\
  \mathcal{A}&=\{ A_1,\cdots,A_M\}.
\end{align}
 Let $\rho$ denote the reduced density matrix of the system on $A$. We define $C(N,M,q)$ as the average of $D_M(\rho,\mathcal{A})$ over different random labelings of the qubits,
\begin{align}\label{eq_C}
  C(N,M,q)=\mathbb{E}\qty[D_M(\rho,\mathcal{A})].
\end{align}
Roughly speaking $C(N,M,q)$ measures irreducible multipartite correlations between $M$ randomly chosen subsystems, each of size $\sim qN$. In this work, we study the behavior of $C(N,M,q)$  as one changes $q$ and $M$ (the number of parties), and discuss how it can be used as an order parameter for k-local nontrivial states.

The basic idea is that the presence of generic irreducible multipartite entanglement for a macroscopic number of parties in a quantum state suggests that the state is non-trivial. To gain a better understanding of this fact, it is helpful to see what can be said about $C(N,M,q)$ on general grounds, before considering specific examples.

 Consider evaluating $C(N,M,q)$ for a state which belongs to a particular k-local nontrivial state. The set $A$ in Eq.\eqref{eq_A} can be thought of as the random subset which includes each qubit with probability $p=Mq$. In the small sampling rate limit $p\simeq 0$, we expect $A$ to be a set of sparse uncorrelated qubits and hence $\rho$ (the reduced density matrix on $A$), to be indistinguishable from a typical state in the trivial phase, which in turn implies we can not distinguish the underlying phase of matter from the trivial phase. However, for large sampling rates $p \simeq 1$, we expect the global state to be approximately recoverable from $\rho$ as we have access to global information, and hence, we can determine the underlying phase of matter.  This observation suggests a phase transition in the complexity of the density matrix of a random subsystem~\cite{hastings} as a function of sampling rate $p$.  The quantity $C(N,M,q)$ is designed to capture this phase transition.

To see this, consider fixing $M\gg 1$ and changing $q$ from $0$ to $1/M$. For $p=Mq$ near zero, as mentioned before, the density matrix $\rho$ is approximately a tensor product of independent density matrices which implies $C(N,M,q)\approx 0$. As we increase $q$, $\rho$ starts to see some correlations between few $A_j$s, but since $D_M$ catches only \textit{irreducible} M-partite correlations and since $M\gg 1$, we still get $C(N,M,q)\approx 0$. However, as soon as $p=Mq$ becomes large enough that the global entanglement structure of the state can be read off from $\rho$, $C$ becomes non-zero. On the other hand, as we increase $q$ further, the global information would now become accessible by the marginals of $\rho$ as well and therefore $C$ drops to zero again. This discussion suggests two phase transitions happen as one tunes $q$ from $0$ to $1/M$; first, when $\rho$ starts to see the global entanglement at $Mq=p_c$, and the second, when this information becomes accessible to the marginals of $\rho$ as well, i.e. at $(M-1)q=p_c$.

We will back up the scenario described in the previous paragraph by (1) Computing $C$ in various specific examples in the next section (random stabilizer states and the toric code) and Appendices  \ref{apx_expander_review}, \ref{holest} (quantum expander codes and a holographic state). (2) Providing a geometrical picture in terms of bond percolation in the case of geometric topological phases. (3) Providing a more quantitative argument in the case of error correcting codes with a finite erasure threshold in Section~\ref{secerase}.

\section{Examples}\label{secex}
\subsection{Random Haar and random stabilizer states}

In this subsection, we study the behavior of $C(N,M,q)$ for random Haar and random stabilizer states. The entanglement structure of subsystems of random Haar and random stabilizer states is well studied. In particular, if $\ket{\psi}$ is a random state of $N$ qubits and if $\rho$ denotes its reduced density matrix on a subset $A$ of qubits with size $n_A< N/2$, it is known that the average entanglement entropy of $\rho$ is exponentially close to its maximum possible value, i.e. $n_A$\cite{hayden2006,stabee}:
\begin{align}\label{eq_randomee}
  \mathbb{E}[S(\rho)]= n_A-O(2^{-2(N/2-n_A)}).
\end{align}
As we shall explain below, it follows then that $C(N,M,q)=0$ for $Mq<1/2$ as well as for $(M-2)q>1/2$, up to exponentially small corrections in $N$.

First consider the $Mq<1/2$ case. Let $A$ be the random subset of qubits as defined in Eq.\eqref{eq_A} and let $\rho$ denote the corresponding reduced density matrix. Note that $n_A/N\simeq Mq$ and that the equality becomes exact in the $N\to\infty$ limit. This shows $\mathbb{E}[S(\rho)]=n_A-O(2^{-N(1/2-Mq)})$. Since for any density matrix $\tilde \rho$ on $n_A$ qubits we have $S(\tilde \rho)\le n_A$,
it follows from Eq.\eqref{pk} that $C(N,M,q)=O(2^{-N(1/2-Mq)})$.

For the $(M-2)q>1/2$ case, let $B=\cup_{j=1}^{M-2} A_i,C=A_{M-1},D=A_M$ denote three non-overlapping subsets of $A$ such that $A=B\cup C \cup D$. Note that $|B|/N\simeq (M-2)q>1/2$ and that the equality becomes exact in the $N\to\infty$ limit. Roughly speaking, in this case $B,C,D$ form an approximate quantum Markov chain, and therefore one could recover the whole state of $A$ just from its marginals\cite{hayden2004,fawzi}. To get a more precise formulation, note that $B,C,D$ satisfy the condition of the strong subadditivity bound in Eq.\eqref{bound}. The value of the upper bound can be found using Eq.\eqref{eq_randomee},
\begin{align}
  I(C:D|B)&=S(BC)+S(BD)-S(B)-S(BCD)\nonumber\\
          &=O(2^{-\varepsilon N}).
\end{align}
 Hence we get $C(N,M,q)=O(2^{-\varepsilon N})$.

 \begin{figure}\label{randomfig}
  \includegraphics[width=1.05\columnwidth]{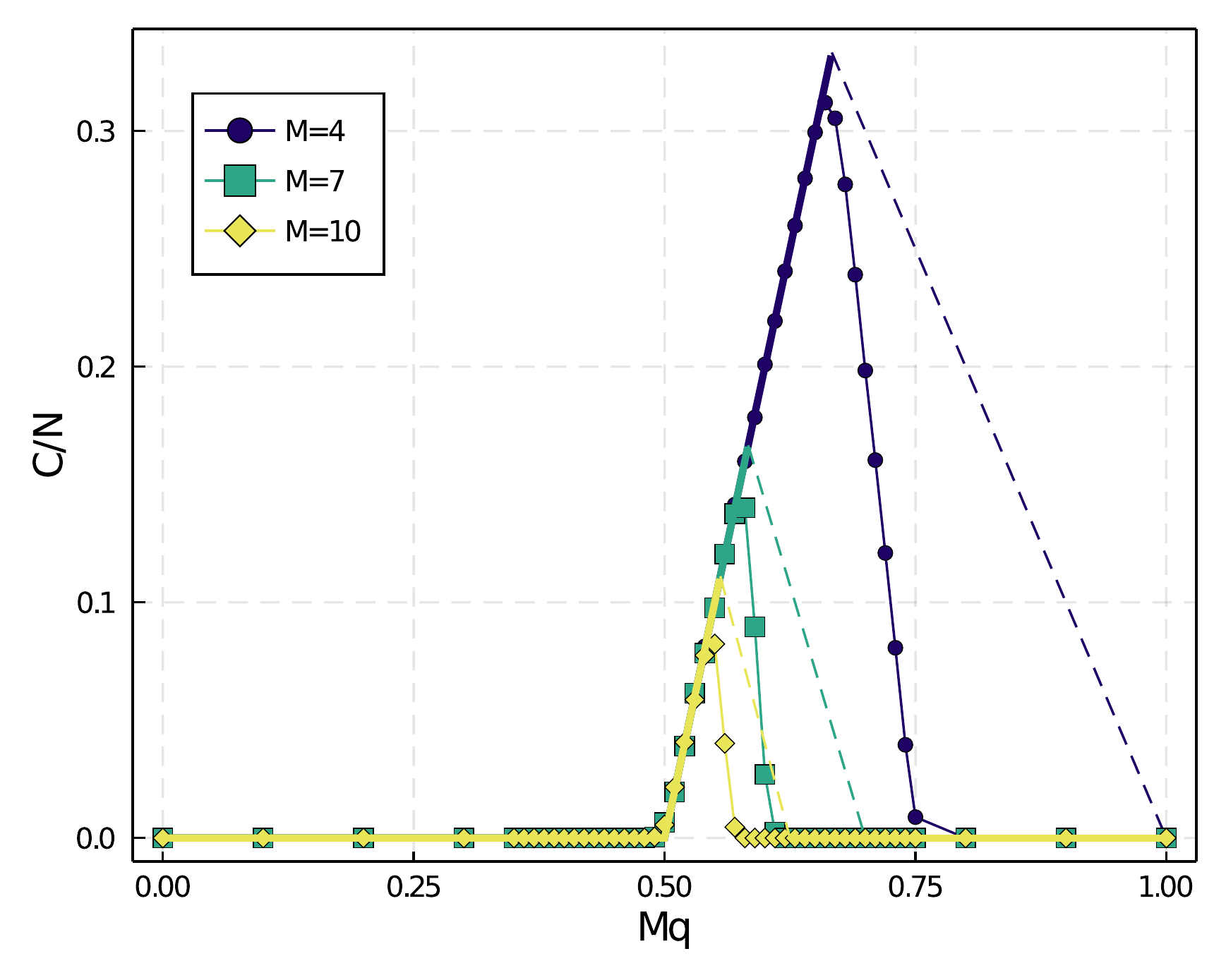}
  \caption{$C(N,M,q)/N$ evaluated numerically for a random stabilizer state, on a fixed system size $N=5184$ and for different $M$'s as a function of $p=Mq$. The thick solid lines correspond to the analytic result. The dashed lines correspond to the analytic upper bound in Eq.\eqref{eq_analytic}}
\end{figure}

We now turn to the regime where $Mq>1/2$ but $(M-1)q<1/2$. First we consider the random stabilizer states. Since $(M-1)q<1/2$ it follows that for large enough $N$, $|A\setminus A_j|/N<1/2$ for any $j=1,\cdots,M$.
Therefore, comparing Eq.\eqref{eq_randomee} and Eq.\eqref{eq_stabee} shows $\mathbb{E}[\dim G_{A\setminus A_j}]$ is exponentially small, which in turn means the contribution of $\mathbb{E}[\dim G_{A;(M-1)}]$
in Eq.\eqref{pk} is exponentially small. On the other hand, $Mq>1/2$ implies that $|A^c|/N<1/2$, where $A^c$ stands for the complement of $A$. Therefore we can find $\dim G_A$ from Eq.\eqref{eq_stabee} by noting $S(A)=S(A^c)$ and using Eq.\eqref{eq_randomee} to find $S(A^c)$. Hence, we find $\mathbb{E}[\dim G_A]/N\simeq 2Mq-1$,
where the equality becomes exact in the thermodynamic limit. This shows $C(N,M,q)/N=2Mq-1$ up to small errors vanishing in the thermodynamic limit. We expect the same behavior in random Haar states because in the intermediate regime, all $M-1$ marginal density matrices are exponentially close to the maximally mixed state and therefore the maximum entropy state consistent with all of them is expected to have the maximum entropy allowed by that system size, i.e. $S(\tilde{\rho})\simeq NMq$, which implies $C(N,M,q)/N\simeq 2Mq-1$, similar to the random stabilizer states.

In the final regime where $(M-1)q>1/2$ but $(M-2)q<1/2$, we can only upper bound $C(N,M,q)$. Note that the subadditivity bound Eq.\eqref{bound} with the choice of regions $B=\cup_{j=1}^{M-2} A_i,C=A_{M-1},D=A_M$ gives (following the same logic as the $(M-2)q>1/2$ regime),
\begin{align}
  I(C:D|B)&=S(BC)+S(BD)-S(B)-S(BCD)\nonumber\\
          &=N-2n_{B}+O(2^{-\varepsilon N}).
\end{align}
This implies that, in the regime $(M-1)q>1/2$ but $(M-2)q<1/2$, we have $ C(N,M,q)/N < 1-2(M-2)q$.

We remark that the combination of this bound with the results in the regime $Mq>1/2,\text{and},(M-1)q<1/2$, already implies that $C(N,M,q)/N$ is non-analytic at $(M-1)q=p_c=1/2$ (this is the second transition discussed in Section \ref{opm}). Numerically, we find that the bound above is not tight in this regime.

Putting everything together, for the random stabilizer state we have the following result:
\begin{align}\
\begin{cases}
C/N=0,&Mq<\frac{1}{2}\\
C/N=2Mq-1,&(M-1)q<\frac{1}{2}<Mq\\
C/N<1-2(M-2)q,&(M-2)q<\frac{1}{2}<(M-1)q\\
C/N=0,              & 1/2<(M-2)q
\end{cases}\label{eq_analytic}
\end{align}
up to small corrections vanishing in large $N$.

Our numerical, as well as analytical results, are displayed in Fig.\ref{randomfig}. The thick solid lines correspond to the analytic expressions and the dashed lines represent the analytical upper bound. The numerical result are computed for a system size $N=5184$. The slight discrepancy between the numerical and analytical results is due to finite size effects and will vanish in the thermodynamic limit\cite{fn_fixedfraction}.  Consistent with the discussion of Sec.\ref{opm}, we find (at least) two phase transitions where the behavior of $C(N,M,q)$ is non-analytic - One where the set of all $M$ parties start to have access to global information about the state $Mq=p_c=1/2$ and a second one where the set of $M-1$ parties starts to have access to global information $(M-1)q=p_c=1/2$.

In analogy with the usual theory of critical phenomena, close to the phase transition at $Mq=1/2$, we find the scaling ansatz
\begin{align}\label{scaling}
&\frac{C(N,M,q=p/M)}{N}= (M-1)^{-\beta} f((p-p_{c})(M-1)^{\alpha}),
 \end{align}
 will collapse the data perfectly with $p_c=1/2$, $\alpha=1$, and $\beta=1$.  In the following subsection, as well as in Appendices \ref{apx_expander_review}, \ref{holest}, we successfully apply the same scaling ansatz to study several analogous phase transitions.

\subsection{Topological Phases of Matter}\label{tpmm}
We now consider the behavior of $C(N,M,q)$ in the conventional topological phases of matter. We will focus on the toric code model, but we expect a similar result to hold for all topological phases.

Consider the toric code model on a $L\times L$ torus. In Sec.\ref{dstab} we discussed the behavior of $D_M$ for a particular choice of subsystems in the toric code. Here we discuss the behavior of $D_M$ for generic subsystems. Let $\mathcal{A}=\{ A_i\}_{i=1}^M $ denote a set of disjoint subsystems of the qubits on the torus. Their union $A=\cup_{i=1}^M A_i$, can be viewed as a collection of patches with some holes on the torus. Let $\rho$ denote the reduced density matrix of the system on $A$.
Stabilizers that stabilize $\rho$ can be taken to be either (1) normal $4$-qubit toric code stabilizers, or (2) stabilizer loops around the holes in $A$, or (3) logical loops around the torus. If we take $M>4$, $D_M(\rho,\mathcal{A})$ can only have contributions from the stabilizers of type (2) and (3). Moreover, such stabilizers contribute only if they are not stabilizing any of its marginals. Hence, we see that $D_M(\rho,\mathcal{A})$ counts the number of independent non-trivial loops in $A$ that are not present in any of its $(M-1)$-party subsystems.  In the context of more general CSS quantum codes, this can be put more formally as a question about how much of the homology/cohomology properties of $A$, thought of as cellulation of a manifold or more generally a chain complex can be read-off from its subsystems.

When the $A_i$ subsystems are chosen randomly as described in section \ref{opm}, one can gain further insights into the behavior of $D_M$ from the theory of bond percolation.
We can view $A$ as the set of qubits on the torus each chosen with probability $p=Mq$. Similarly, any $(M-1)$-party union set, i.e. $A\setminus A_{j}$ for any $j$, can be viewed as the set of qubits on the torus each chosen with probability $p'=(M-1)q$.
Now if $p>p_c$ while $p'<p_c$ (with $p_c=1/2$ denoting the bond percolation threshold on the square lattice), the set $A$ percolates with probability one while no $(M-1)$-party union set does.  Hence, while the two logical qubits are accessible to $A$, their value can not be deduced from any of the $(M-1)$-party marginals.
This ensures that $C(N,M,q)\ge 2$ for $\frac{p_c}{M}<q<\frac{p_c}{(M-1)}$. Note that this region of $q$ shrinks as $M$ increases, which implies that for large $M$s we are close to the percolation critical point and should therefore be careful with the order of limits and finite-size effects.  Proximity to the percolation threshold $Mq\approx p_c$, in turn, implies that $A$ is likely to include macroscopically large holes. Each one of these large punctures (with perimeter larger than $M$) has some chance of contributing to $D_M(\rho,\mathcal{A})$ depending on whether it is present in $M-1$ partite systems or not (note that these contributions are also present on surfaces without a logical qubit, e.g. a sphere).

Putting everything together, we expect $C(N,M,q)$ to be small unless we choose our parameters such that $A$ and its $(M-1)$-party subsystems are on different percolation phases $\frac{p_c}{M}<q<\frac{p_c}{(M-1)}$. Furthermore, the calculation of $C(N,M,q)$ can be mapped into a calculation of the distribution of large punctures in the percolation theory near the critical point. However, we do not attempt this analytical computation here and instead use numerical results to find a phenomenological form for $C(N,M,q)$ and discuss its universal properties which should be entirely set by the percolation CFT.  We expect a similar picture to hold in all topological phases (with stabilizers replaced by generic Wilson loop operators).

Before proceeding further, we'd like to emphasize that the discussion above was simplified by the fact that both $X$ and $Z$ stabilizers of the toric code live on a square lattice (square lattice is self dual). In the case of more generic lattices, or more generic CSS codes, one needs to carefully separate the discussion of $X$ and $Z$ stabilizers to be precise. See Appendix. \ref{hextoric} for a brief discussion of toric code on a hexagonal lattice.

\begin{figure}
  \includegraphics[width=\columnwidth]{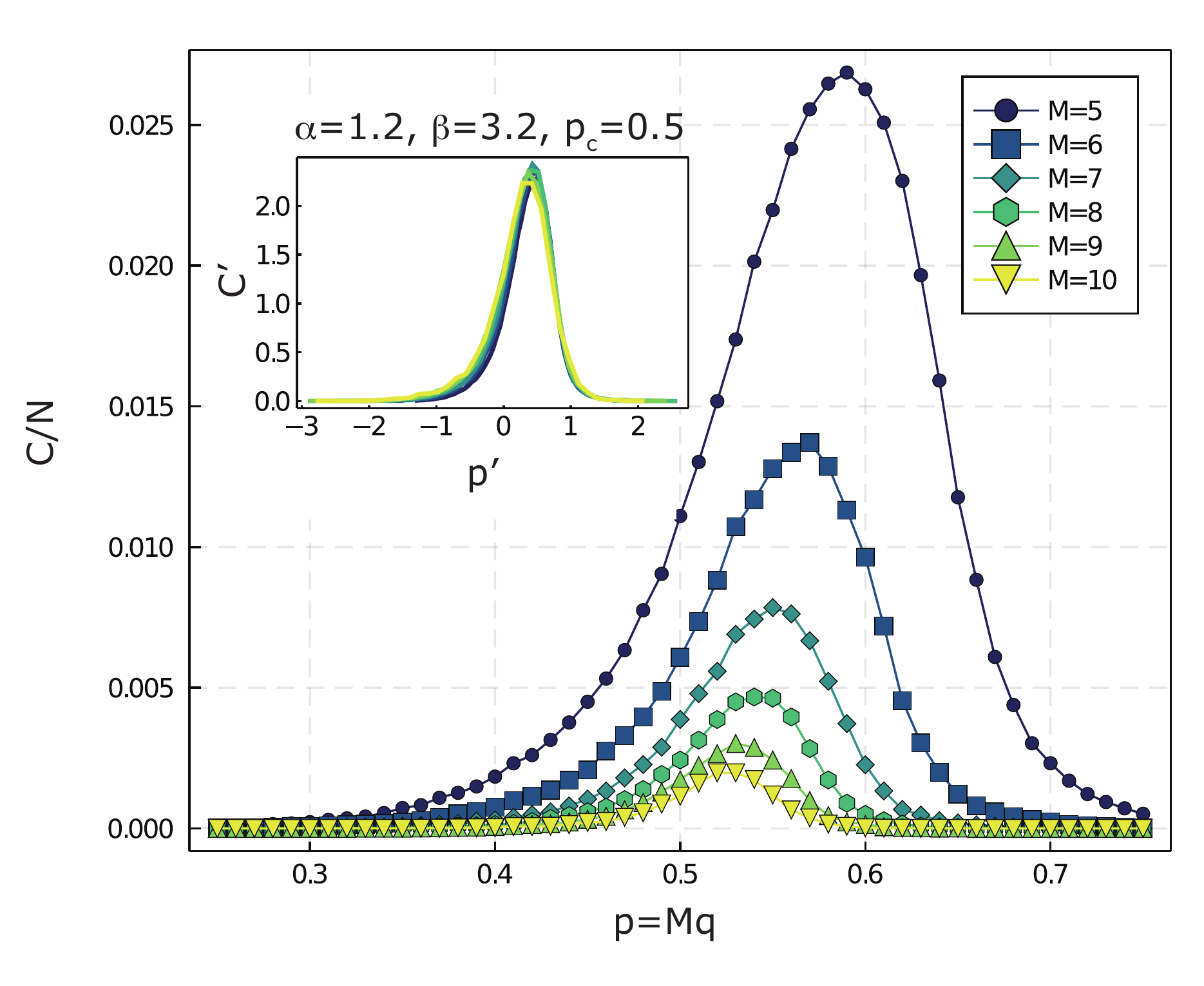}
  \caption{$C(N,M,q)/N$ for the toric code on a torus. Plotted at a fixed system size $N=72\times72$  and for different $M$s as a function of $p=Mq$. Inset: the best data collapse for the toric code around the point $Mq=1/2$, using the scaling ansatz of Eq.\eqref{scaling} with parameters $p_c=0.5$, $\alpha=1.2$ and $\beta=3.2$.}
  \label{trpk}
\end{figure}

We now proceed to present the numerical results for the case of the toric code on a torus (square lattice): Based on numerics, we find that in the large system size limit $N\gg 1$, $C(N,M,q)$ becomes proportional to the system size $N$. This is consistent with the percolation picture because the distribution of punctures is proportional to the area of the system.  In Fig.\ref{trpk}, we have plotted $C(N,M,q)/N$ at fixed system size $N=72\times72$ as a function of $p=Mq$ for a number of different $M$s. Consistent with our expectation, we find that $C(N,M,q)$ is nonzero in a region of width $\Delta p \approx 1/M$ around $p_c=1/2$. Furthermore, the inset shows that we can collapse the data around the $Mq=1/2$ point, using the scaling ansatz of Eq.\eqref{scaling}. We find that $p_c=0.5$, $\alpha=1.2$ and $\beta=3.2$ results in the best collapse.

\subsection{Using $C(N,M,q)$ to detect a phase transition}

To claim that the behavior of $C(N,M,q)$ can be used to distinguish trivial and non-trivial states we need to understand how $C(N,M,q)$ behaves across a phase transition. As an example, we look into the phase transition from the toric code fixed point into the trivial phase. In the Hamiltonian setting, this can be done by following the ground state of the troic code in an external magnetic field~\cite{zeng2019quantum}. This approach is not suitable for us since our computational power in computing is $C(N,M,q)$ is limited to stabilizer states. It has been recently realized that in the context of random hybrid quantum circuits we can stabilize topological phases and study their transitions entirely within the stabilizer formalism~\cite{lava1,lava2}. This convenient feature makes these models suitable to study using our computational method.
\begin{figure}
  \includegraphics[width=\columnwidth]{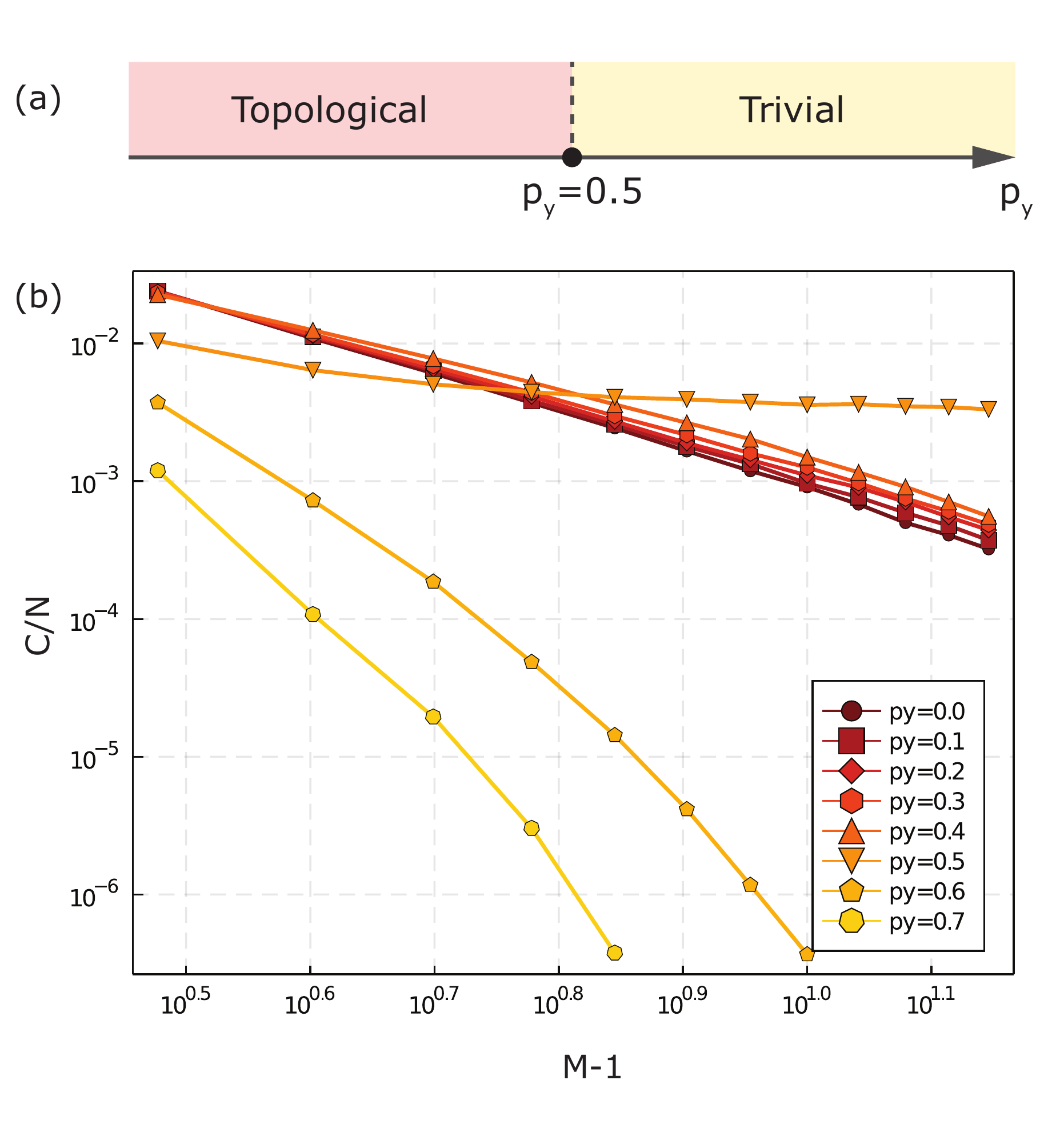}
  \caption{a) The phase diagram of the random quantum circuit, consisted of measuring either single qubit $Y$ with probability $p_y$ or four qubit toric code stabilizers with probability $1-p_y$. b) $C(N,M,q)$ at $Mq=p_c=0.5$ and $N=56\times 56$ as a function of $M$ evaluated on the late time states of the random quantum circuit and average over many realizations.}
  \label{fig_trph}
\end{figure}

 In particular, we shall use the model developed in Ref.~\cite{lava2}. We consider a random quantum circuit starting with an arbitrary state. At each step of the circuit, with probability $p_y$ a randomly chosen single qubit is measured in the $Y$ basis, and with probability $1-p_y$ a random stabilizer of the toric code is measured. The entanglement structure of the late time state of this circuit exhibits an entanglement phase transition at $p_y=1/2$, as shown in Fig.\ref{fig_trph}a. We refer the reader to Ref.~\cite{lava2}, for further details.

 We study the behavior of the $C(N,M,q)$ averaged over different circuit realizations as we change $p_y$ across the phase transition. In particular, for each $p_y$,  we look at $C(N,M,q)$ at $Mq=0.5$ as a function $M$. Based on the ansatz in Eq.\eqref{scaling}, we expect $C(N,M,q=p_c/M)$ to be a power law as a function of $M-1$ in the topological phase. On the other hand, in the trivial phase, we expect $C$ to vanish faster than any power-law due to the absence of long-range entanglement. The numerical results are presented in Fig.\ref{fig_trph}b.
As evident from the plot, $C(N,M,q=p_c/M)/N$ behaves as a power-law throughout the topological phase. The exponent also seems to remain constant (within the error bar), suggesting that the power-law shape of $C(N,M,q=\frac{1/2}{M})$ as well its exponent are universal in the topological phase. In the trivial phase $p_y>0.5$, we find that $C(N,M,q=\frac{1/2}{M})$ decays faster than any power law (shown in Fig.\ref{fig_trph}b) - establishing a sharp, qualitative difference between the behavior $C(N,M,q=p_c/M)$  in trivial and topological phases.

\vspace{1cm}
Before ending this section we remind the reader that two additional examples, i.e. quantum expander codes and holographic are studied and discussed in Appendices \ref{apx_expander_review}, \ref{holest}.

 \section{Relation to erasure threshold}\label{secerase}

 In this subsection, we discuss the relation between $C(N,M,q)$  and the erasure threshold in quantum error correcting codes.
 Let $\rho$ be a given code state in a quantum error correcting code $\mathcal{Q}$. Consider the situation in which the subset $E$ of the qubits are erased. The quantum error correcting code $\mathcal{Q}$ can correct for this erasure error if and only if it can reconstruct $\mathcal{\rho}$ using only the remaining qubits. We say an error correcting code has a finite erasure threshold $e_\text{th}>0$ if with probability one it can correct random erasure errors below a certain rate $e<e_\text{th}$, where $e$ is the probability of a given qubit being erased. Equivalently, in a quantum error correcting code with erasure threshold $e_\text{th}$, if we consider a subset $A$ of qubits where each qubit is kept at random with probability $p>(1-e_\text{th})$, it includes all the information about the state of the logical qubits with probability one. By no-cloning theorem, this implies that the erased subsystem includes no information about the state of the logical qubits.

Consider the stabilizer state $\rho=\ketbra{0_L}$ in a stabilizer code $\mathcal{Q}$ with a finite erasure threshold $e_\text{th}$. Let $Z_L$ denote the logical operator associated with $\ket{0_L}$, which along with other stabilizers, generates the stabilizer group of $\rho$. Consider the closely related stabilizer state $\tilde{\rho}=\frac{1}{2}(\ketbra{0_L}+\ketbra{1_L})$. The generators for the stabilizer group of $\tilde{\rho}$ can be taken to be the same as the generators of the stabilizer group for $\rho$ but with $Z_L$ dropped.
For a given $M$ and $q$, consider $A$ to be the randomly chosen subsystem defined in section \ref{opm} and let $\rho_A=\tr_{A^c}(\rho)$ and $\tilde{\rho}_A=\tr_{A^c}(\tilde{\rho})$ denote the associated reduced density matrices on $A$. From the discussion in the previous paragraph, we know that if $Mq<e_{th}$, with probability one $\rho_A$ does not include $Z_L$ or a logically equivalent operator (for brevity, in the rest of this paragraph we drop the phrase ``or a logically equivalent operator"). We also know that if $Mq>1-e_{th}$, with probability one $\rho_A$ does include $Z_L$. Now consider the smallest $Mq=p_c$ such that $\rho_A$ has a nonzero probability to include $Z_L$. Note that at this point, marginals of $\rho_A$ do not include $Z_L$, therefore, all marginals of $\rho_A$ and $\tilde{\rho}_A$ would be equal. In that case, the stabilizer group associated with $\rho_A$ has a finite chance to have one extra generator which is not present in the stabilizer group of $\tilde{\rho}_A$, which in turn entails that there is a finite chance to get $S(\tilde{\rho})>S(\rho_A)$
(see Eq.\eqref{eq_stabee}). Therefore we are guaranteed to get $C(N,M,q)>0$ for the interval $(M-1)q< p_c< Mq$ (Note that $e_{th}<p_c<1-e_{th}$). Roughly speaking, this implies that all states which belong to an error correcting code with a finite erasure threshold have long-range irreducible multipartite correlations amongst their random subsystems (for suitably chosen subsystem sizes).

\section{Final remarks and future directions}\label{secdics}
In this work, we proposed a non-geometric order parameter for k-local nontrivial phases and demonstrated its utility by studying a wide variety of examples. In this section, we conclude by speculating about the behavior of $C(N,M,q)$ in generic quantum systems and list several interesting future directions.

 The following is a summary of the expected behavior of  $C(N,M,q)$  in generic systems: (1) $C(N,M,q)$ is expected to be nonzero in all geometric topological phases, including states defined on surfaces without a ground state degeneracy, e.g. the surface of a sphere. This is because even though the contribution from the logical qubit vanishes, the large holes close to percolation transition still contribute to $C(N,M,q)$ (see Section \ref{secex}). Therefore, a nonzero value of $C(N,M,q)$ is not necessarily tied with the existence of a protected logical qubit. (2) $C(N,M,q)$ is expected to be nonzero in all ``k-local" quantum error-correcting codes with a finite erasure threshold (see Section \ref{secerase}). (3) $C(N,M,q)$ is expected to be zero in some gapless phases. In particular, $C(N,M,q)$ is expected to vanish in eigenstates of free systems. This is because in these systems, Wick's theorem implies that two partite density matrices include all the global information. i.e. $C(N,M>2,q)=0$. (4) $C(N,M,q)$ is expected to be nonzero in gapless phases that have error correcting properties. Examples include the Motzkin chain~\cite{movassagh2016supercritical,Brandao} and CFTs important in AdS/CFT correspondence~\cite{almheiri2015bulk,pastawski2015holographic}. It would be intriguing to understand the behavior of $C$ in these systems.

Our work has demonstrated the basic utility of $C(N,M,q)$ in studying k-local systems - However, much work remains to be done. In particular: (1) It would be interesting if the computation scheme used in this paper can be extended to states beyond the stabilizer states. (2) While we demonstrated, the applicability of the scaling ansatz in Eq.\eqref{scaling} to many examples, we do not have a clear understanding of the meaning of these exponents and the underlying field theory. In particular, it is plausible that the value of these critical exponents encodes some universal information (e.g. dimensionality) of the underlying k-local phase. (3) It would be interesting if we can understand the relation between $C(N,M,q)$ and the circuit complexity. In particular, it is intriguing to see if it is possible to prove the non-triviality of a state that has power-law decaying $C(N,M,q)$ (as a function of $M$). (4) It is interesting to understand if our order parameter can be used to differentiate distinct volume law phases.

\section*{acknowledgements}
We thank John McGreevy, Tarun Grover, Maissam Barkeshli, and Bowen Shi for useful discussions. The authors acknowledge the University of Maryland supercomputing resources (http://hpcc.umd.edu) made available for conducting the research reported in this paper. A.L is supported by NSF CAREER (DMR- 1753240), Alfred P. Sloan Research Fellowship, and JQI- PFC-UMD. Y.A is supported by Simons Collaboration on UltraQuantum Matter, grant 651440 from the Simons Foundation as well as University of California Laboratory Fees Research Program, grant LFR-20-653926
\bibliographystyle{apsrev4-1.bst}
\bibliography{library.bib}

\begin{thebibliography}{34}
\bibitem [{fn_()}]{fn_fixedfraction}%
  \BibitemOpen
  \href@noop {} {\ }\bibinfo {note} {This can be numerically verified by
  considering a slightly different way of choosing $A_i$ subsystems which
  converges to the thermodynamic limit much faster: instead of assigning iid
  labels to each qubit which results in $|A_i|/N=q+O(1/\sqrt{N})$, assign
  distinct random sets of $\lfloor qN\rfloor$ qubits to each set
  $A_i$}\BibitemShut {NoStop}%
\end{thebibliography}

\pagebreak
\clearpage
\appendix

\section{Quantum Expander Code}\label{apx_expander_review}
\subsection{Review of classical and quantum expander codes}
\textit{Classical linear codes -}
A classical linear code $\mathcal{C}$ over $n$ bits can be defined as the kernel of a $m\times n$ parity check matrix $H\in \mathbb{F}_2^{m\times n}$, i.e. $c\in \mathbb{F}_2^n$ is a code word in $\mathcal{C}$ iff $Hc=0$. A classical code associated with the parity check matrix $H$ encodes $k=\dim(\ker(H))=n-\rank(H)$ logical bits and has distance $d$ where $d$ is the minimum Hamming distance between pairs of different code words in $\mathcal{C}$.
For future convenience, we set $d=\infty$ for codes with $k=0$.
The standard notation $\mathcal{C}=[n,k,d]$ is used to compactly specify the code properties. Note that $k\ge n-m$ since $\rank(H)\le m$. Furthermore, different parity check matrices can give rise to the same classical code $\mathcal{C}$.

\begin{figure}
  \includegraphics[width=\columnwidth]{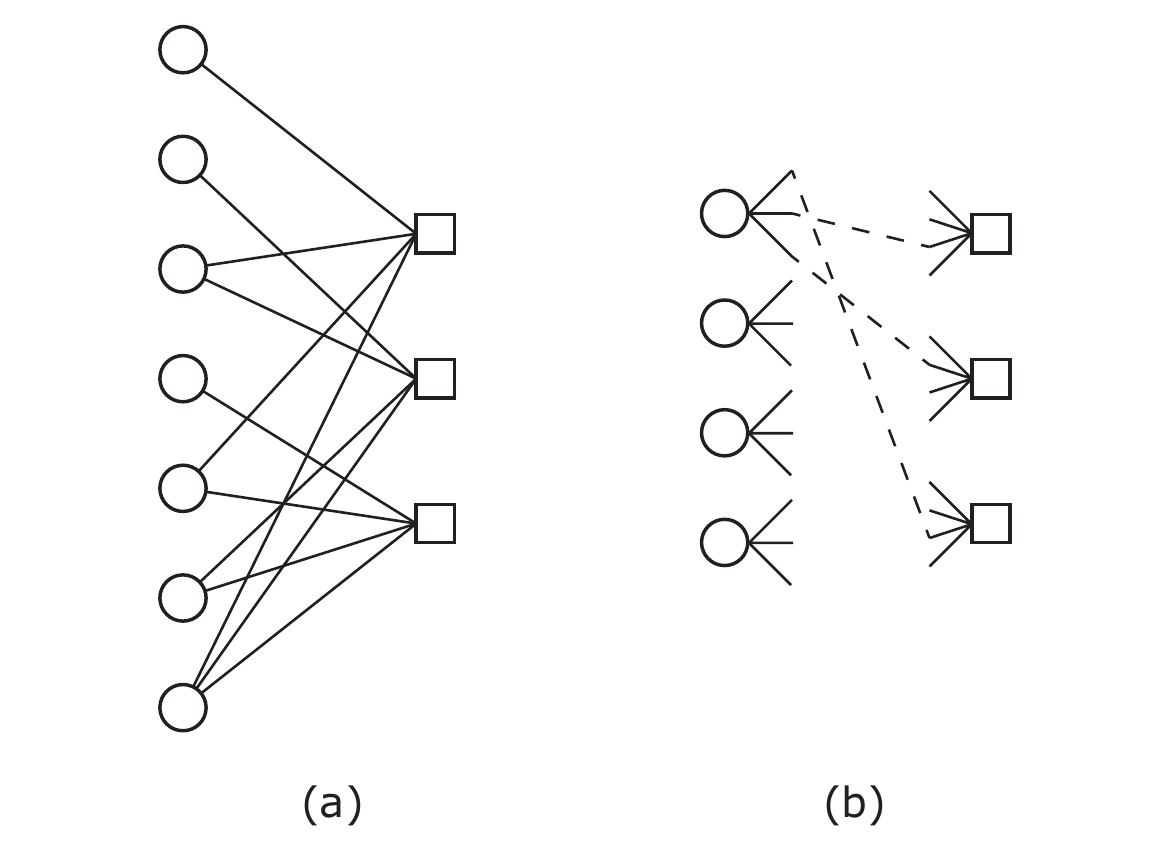}
  \caption{(a) The Tanner graph associated with the $7$-bit Hamming code, given by the parity check matrix in Eq.\eqref{eq_hampar}. (b) Random construction of biregular graphs for $(\Delta_L,\Delta_R)=(3,4)$.}
  \label{fig_tanner}
\end{figure}

A classical code can also be specified by a Tanner graph. For a given classical linear code and its parity check matrix $H\in \mathbb{F}_2^{m\times n}$, its Tanner graph is a bipartite graph with $n$ bit vertices on the left and $m$ check vertices on the right. A check vertex $i$ is connected to a bit vertex $j$ iff $H_{i,j}=1$. Since the parity check matrix of a code is not uniquely defined, so neither is its Tanner graph. As an example, the Tanner graph of the $[7,4,3]$ Hamming code with the parity check matrix of,
\begin{equation}\label{eq_hampar}
H=
\begin{bmatrix}
  1&0&1&0&1&0&1\\
  0&1&1&0&0&1&1\\
  0&0&0&1&1&1&1
\end{bmatrix}
\end{equation}
is shown in Fig.\ref{fig_tanner}a.
Conversely, given a bipartite graph, one can define a parity check matrix $H$ accordingly and associate a classical code $\mathcal{C}$ to it.

For a given classical code $\mathcal{C}=[n,k,d]$ and its parity check matrix $H\in \mathbb{F}_2^{m\times n}$, the transpose code of $\mathcal{C}$, denoted by $\mathcal{C}^T=[m,\tilde k,\tilde d]$ is defined as the classical code associated to the parity check matrix $H^T$ which is the matrix transpose of $H$. Note that $\mathcal{C}^T$ depends on the choice of parity check matrix for $\mathcal{C}$ and is not uniquely defined by $\mathcal{C}$. Since $\rank(H)=\rank(H^T)$,
it follows that $\tilde k=k-(m-n)$. The Tanner graph for $\mathcal{C}^T$, is the same as $\mathcal{C}$ with bit and check vertices interchanged.

\textit{Classical Expander Code -}
Consider a bipartite graph $G=(V,C,E)$ with $V$ and $C$ denoting the set of vertices on left and right respectively and $E$ denoting the set of edges between $V$ and $C$. The degree of a vertex is the number of edges connected to it. A bipartite graph is said to be biregular if all vertices on each part have the same degree. Let $G$ be a biregular graph with left and right degrees $\Delta_L$ and $\Delta_R$ respectively, i.e. all vertices in $V$ have degree $d_L$ and all vertices in $C$ have degree $d_R$. We say $G$ is $(\alpha,\delta)-$left expanding if for any subset of left vertices $S\subset V$ for which $|S|\le \alpha |V|$, we have $|\Gamma(S)|\ge (1-\delta) \Delta_L |S|$ where $\Gamma(S)$ is the set of neighbors of $S$ on the graph $G$.
The classical code whose Tanner graph is $(\alpha,\delta)$-left expanding for some positive $\alpha$ and $\delta<1/2$, is an instance of classical expander code. The distance of such a code is lower bounded by $\alpha n$ \cite{richardson2008modern}.

A biregular graph $G$ with good expansion may be obtained via random construction. Imagine we want to construct a biregular graph $G$ with $n$ vertices of degree $\Delta_L$ on the left and $m=\frac{\Delta_L}{\Delta_R}n$ vertices of degree $\Delta_R$ on the right. We start by putting $n$ vertices on the left with $\Delta_L$ stubs connected to each and putting $m$ vertices on the right, each with $\Delta_R$ stubs connected to it. Note that there are $N=n \Delta_L=m \Delta_R$ stubs on each side. Now, we choose a random permutation $\pi$ of numbers $1,\cdots,N$ uniformly at random and for each $i$, we connect the $i$'th stub on left to the $\pi_i$'th stub on the right (see Fig.\ref{fig_tanner}b). It can be shown that for any $\delta>1/\Delta_L$,
there exists a strictly positive $\alpha$ such that the resulting graph is $(\alpha,\delta)$-expanding with a probability that approaches $1$ for large enough $n$s \cite{richardson2008modern}.

\textit{Hypergraph product code -}
The hyper graph product code is a quantum code built out of two classical codes. Let $\mathcal{C}_1=[n_1,k_1,d_1]$ and $\mathcal{C}_2=[n_2,k_2,d_2]$ denote two classical codes corresponding to two given parity check matrices $H_1\in \mathbb{F}_2^{m_1\times n_1}$ and $H_2\in \mathbb{F}_2^{m_2\times n_2}$ respectively.
The hypergraph product code of $\mathcal{C}_1$ and $\mathcal{C}_2$ is a CSS stabilizer code $\mathcal{Q}$ defined on $n_1n_2+m_1m_2$ qubits. The $X$ and $Z$ stabilizers of the quantum code $\mathcal{Q}$ is given by the parity matrices $H_X$ and $H_Z$ which are defined as
\begin{align}
  H_X&=\qty(I_{n_1}\otimes H_2~|~H_1^T  \otimes I_{m_2})\\
  H_Z&=\qty(H_1\otimes I_{n_2}~|~ I_{m_1}  \otimes H_2^T),
\end{align}
where $I_n$ is the $n$-dimensional identity matrix and $(A~|~B)$ is the matrix which results from the horizontal concatenation of $A$ and $B$ matrices. It can be shown that $\mathcal{Q}$ encodes $k_\mathcal{Q}=k_1 k_2 + \tilde {k}_1 \tilde {k}_1$ logical qubits and its distance is lower bounded by $\min(d_1,d_2,d_1^T,d_2^T)$, or  in the standard notation $$\mathcal{Q}=[\![ n_1n_2+m_1m_2, k_1k_2+k_1^T k_2^T, \min(d_1,d_2,d_1^T,d_2^T) ]\!].$$

\textit{Quantum expander code -}
Quantum expander code is the hypergraph product code of a classical expander code with itself \cite{leverrier2015quantum}.

\subsection{Numerical results for the quantum expander code}
Code states of a $[\![n,k,d ]\!]$ quantum error correcting code with $k>0$ and $d$ growing with system size can not be prepared via constant-depth k-local unitary circuits \cite{pcp,aharonov2018quantum}. Moreover, quantum LDPC codes that have distance scaling as a power law with system size have a finite error threshold \cite{Pryadko}.  Therefore, it is interesting to study the irreducible multipartite entanglement in the quantum states which correspond to code words of a quantum error correcting code.

\begin{figure}
  \includegraphics[width=\columnwidth]{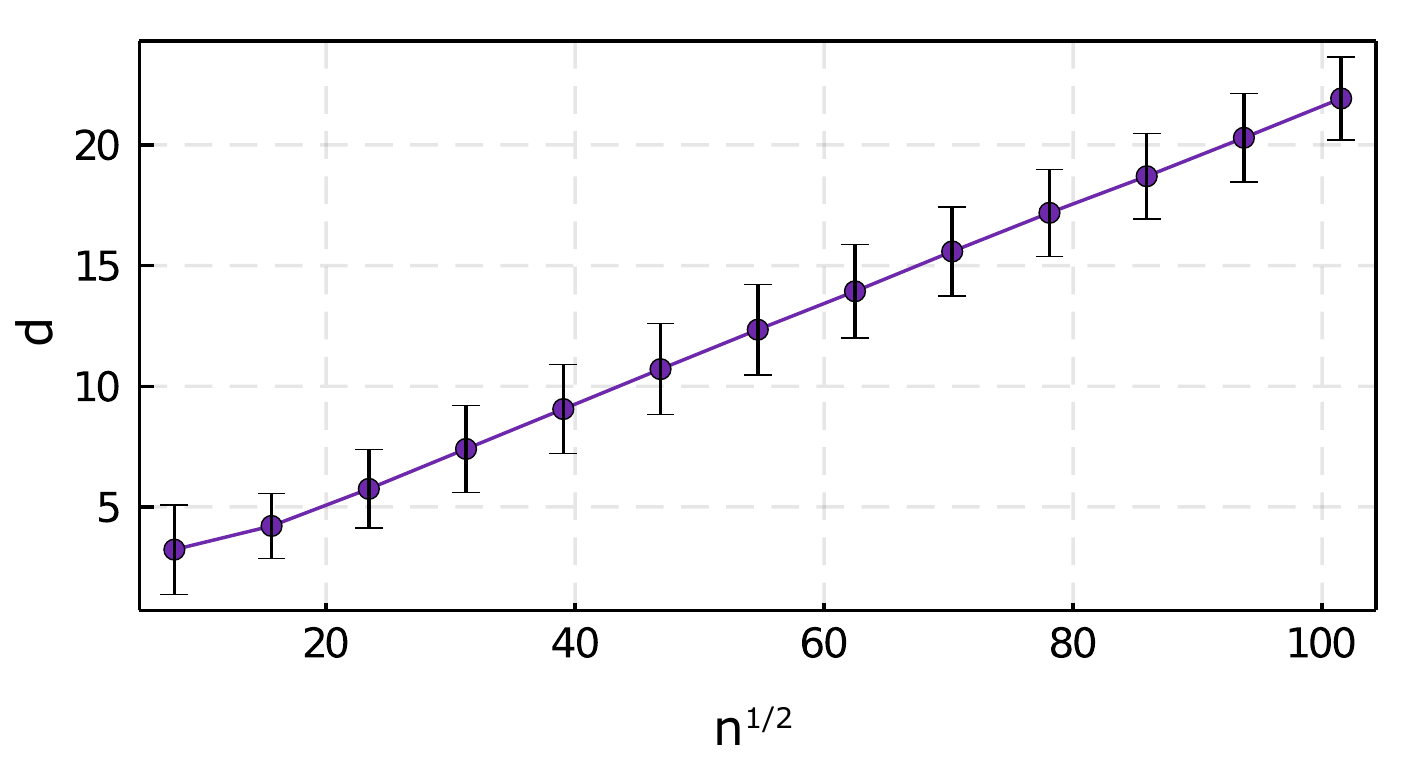}
  \caption{Average distance and its variance of quantum expander code with $n$ qubits as a function of $\sqrt{n}$, based on random bi-partite graphs with left and right degrees $5$ and $6$ respectively.}
  \label{fig_distance}
\end{figure}

In particular, we will focus on the quantum expander code family \cite{leverrier2015quantum}. Quantum expander codes have finite rate and their distance scales as $\sqrt{n}$, so they make a family of $[\![n,k=\Theta(n),d=\Omega(\sqrt{n}) ]\!]$ quantum LDPC codes. In what follows, we consider quantum expander codes on $n$ qubits which are based on classical random bi-partite graphs with left and right degrees $\Delta_L=5$ and $\Delta_R=6$ respectively.
It can be shown \cite{richardson2008modern} that with probability $P>1-O(n^{-\beta})$, the resulting quantum code has distance $d>\alpha \sqrt{n}$ for some positive constants $\alpha$ and $\beta$ and encodes $k=\frac{(\Delta_R-\Delta_L)^2}{\Delta_R^2+\Delta_L^2} n=\frac{1}{61}n$ logical qubits. This result in an upper bound on the erasure threshold of $e_\text{th}\le(1-1/61)/2\simeq 0.49$ \cite{delfosse2012upper}.
Fig.\ref{fig_distance} shows the average code distance of this family and its variance as a function of $\sqrt{n}$.

\begin{figure}
  \includegraphics[width=\columnwidth]{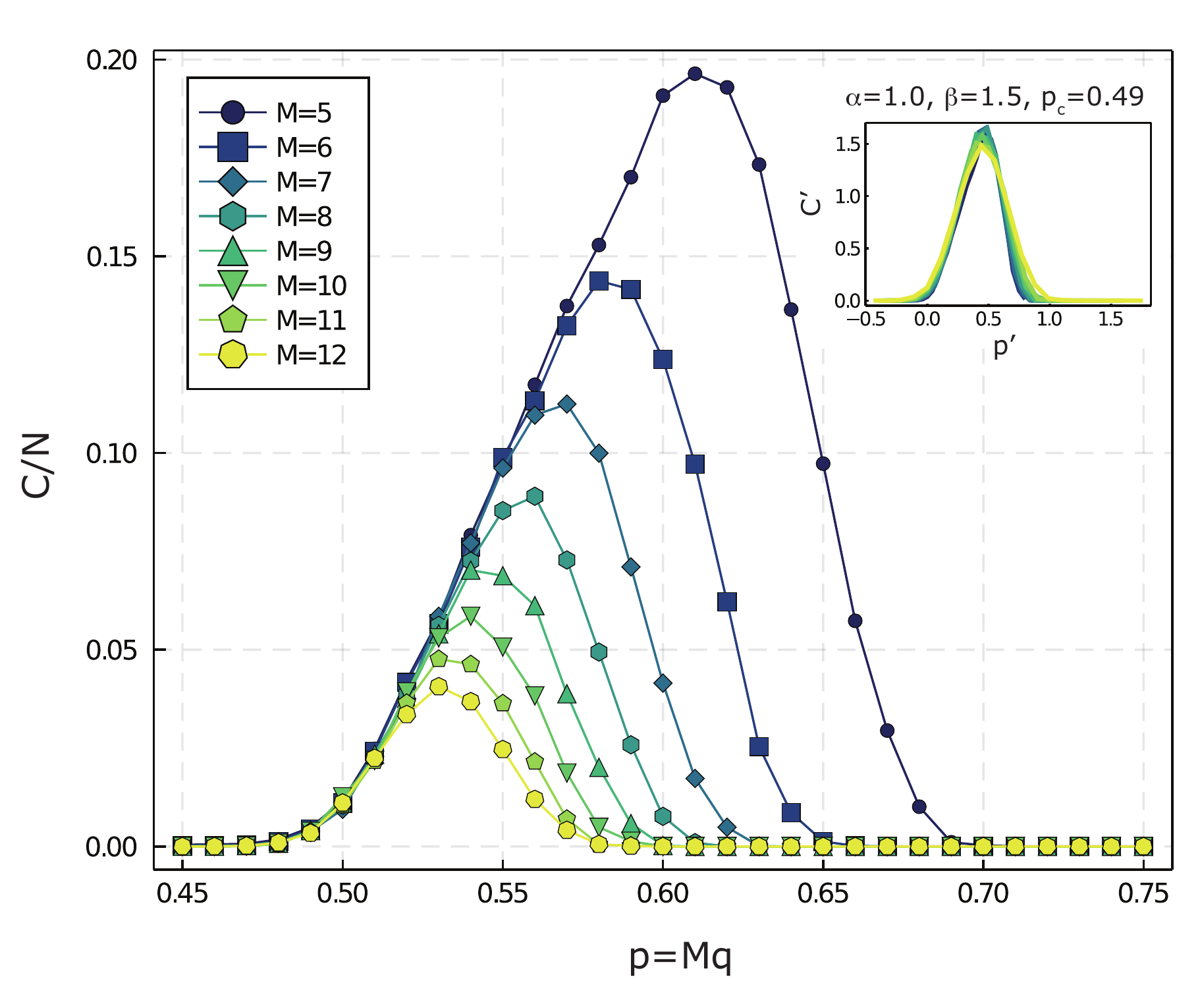}
  \caption{$C(n,M,q)$ as a function of $p=Mq$ for different $M$s at fixed $n=2196$. The inset shows the collapsed data corresponding to the first phase transition, where the horizontal axis is rescaled to $p'=(Mq-p_c)(M-1)^\alpha$ and the vertical axis is rescaled to $C'=C~(M-1)^\beta$, with $p_c=0.49, \alpha=1.0$ and $\beta=1.5$.}
  \label{fig_expander_plots}
\end{figure}

We now present the numerical results corresponding to the irreducible many-partite correlations for code states of the aforementioned quantum expander codes. All results are averaged over different selections of random partition as well as different realizations of random quantum expander codes.  Fig.\ref{fig_expander_plots} shows $C(n,M,q)$ as a function of $p=Mq$ for different values of $M$ at fixed system size of $n=2196$. As expected from general considerations, $C$ vanishes for small and large values of $p$ and becomes nonzero in a window of width $\Delta p \sim \frac{1}{M}$ near $p \sim 1/2$. The inset shows the collapsed data according to the scaling ansatz in Eq.\eqref{scaling}
We find $p_c=0.49$, $\alpha=1.0$ and $\beta=1.5$ collapses different curves on top of each other.

\section{Holographic states}\label{holest}
\begin{figure}
  \includegraphics[width=0.6\columnwidth]{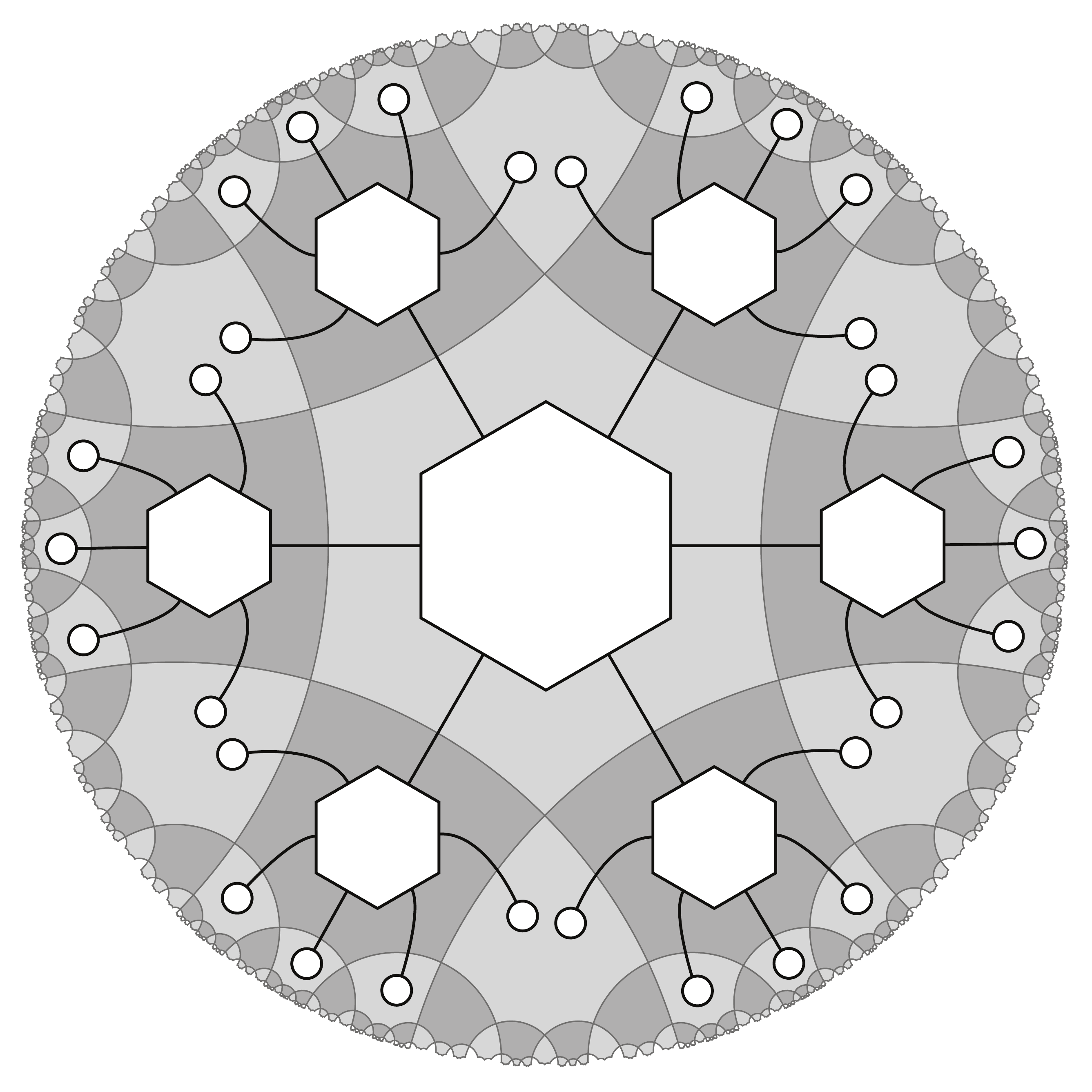}
  \caption{The holographic state on $n=30$ qubits constructed by two layers of $6$-legged perfect tensors. The white dots denote the physics qubits. }
  \label{fig_holostatetensornet}
\end{figure}

\begin{figure}
  \includegraphics[width=\columnwidth]{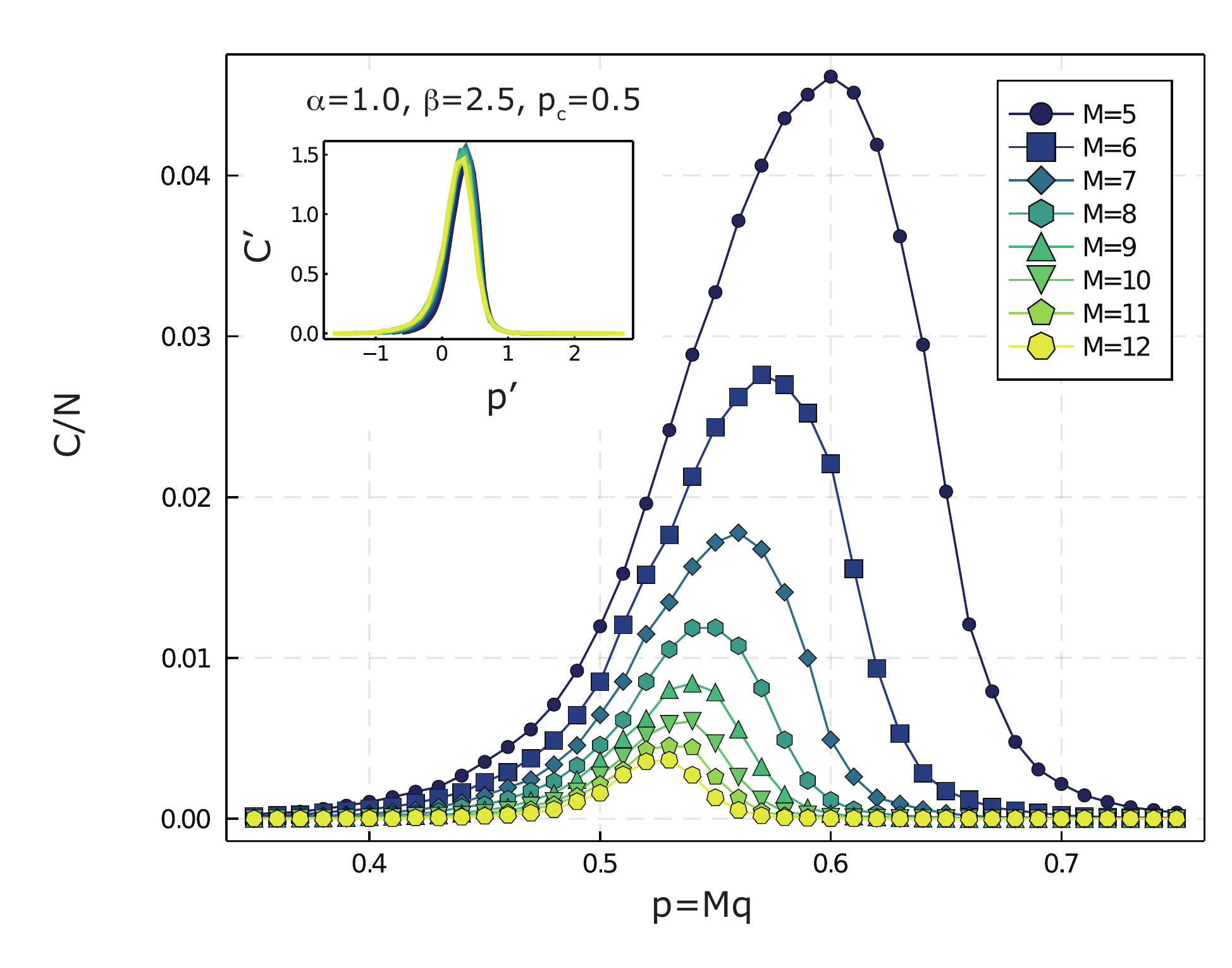}
  \caption{$C(N,M,q)$ as a function of $Mq$ for different $M$s, computed for the the holographic hexagon state with  $N=1590$ physical qubits. The inset shows the data collapse according to the scaling ansatz of Eq.\eqref{scaling}, with  $p_c=0.5$, $\alpha=1.0$ and $\beta=2.5$. The horizontal axis of inset is $p'=(Mq-p_c)(M-1)^\alpha$ and the vertical axis is $C'=(M-1)^\beta C$. }
  \label{fig_holo_plots}
\end{figure}

The holographic quantum error correcting codes and the closely related holographic states were introduced as simple toy models of the AdS/CFT correspondence \cite{pastawski2015holographic, jahn2021holographic}. These constructions utilize a special type of tensor, known as a perfect tensor. The tensor network which results from contracting a finite number of perfect tensors according to a compatible tessellation of the hyperbolic plane represents a holographic quantum error correcting code or a holographic state depending on whether the tensors in the bulk have any uncontracted leg left or not. The dangling legs on the boundary correspond to physical qubits while the uncontracted bulk legs (if any) correspond to logical qubits. Here, we focus on a holographic state which is defined by contracting $6$-legged perfect tensors according to the $\{6,4\}$-tiling of the hyperbolic plane (see Fig. \ref{fig_holostatetensornet}). The perfect tensor in this construction corresponds to a $[\![6,0 ]\!]$ stabilizer code which is closely related to the well-known $5$-qubit code. This makes the resulting holographic state, a stabilizer state as well (see Ref. \cite{pastawski2015holographic} for more details). In what follows we refer to this state as the holographic hexagon state.

Fig.\ref{fig_holo_plots} shows $C(n,M,q=p/M)$ as a function of $p=Mq$, for fixed $n=5934$ and different $M$s. This particular number of $n$ corresponds to contracting $r=6$ layers of edge-adjacent hexagons on the hyperbolic plane (see Fig.\ref{fig_holostatetensornet} for $r=2$).  Similar to other cases, $C$ vanishes for small and large $p$s and it peaks near $p=0.5$. The inset of the plot, shows the same data but collapsed according to the scaling ansatz of Eq.\eqref{scaling},
with parameters $p_c=0.5$, $\alpha=1.0$ and $\beta=2.5$.

\section{Toric code on a hexagonal lattice}\label{hextoric}

\begin{figure}
  \includegraphics[width=\columnwidth]{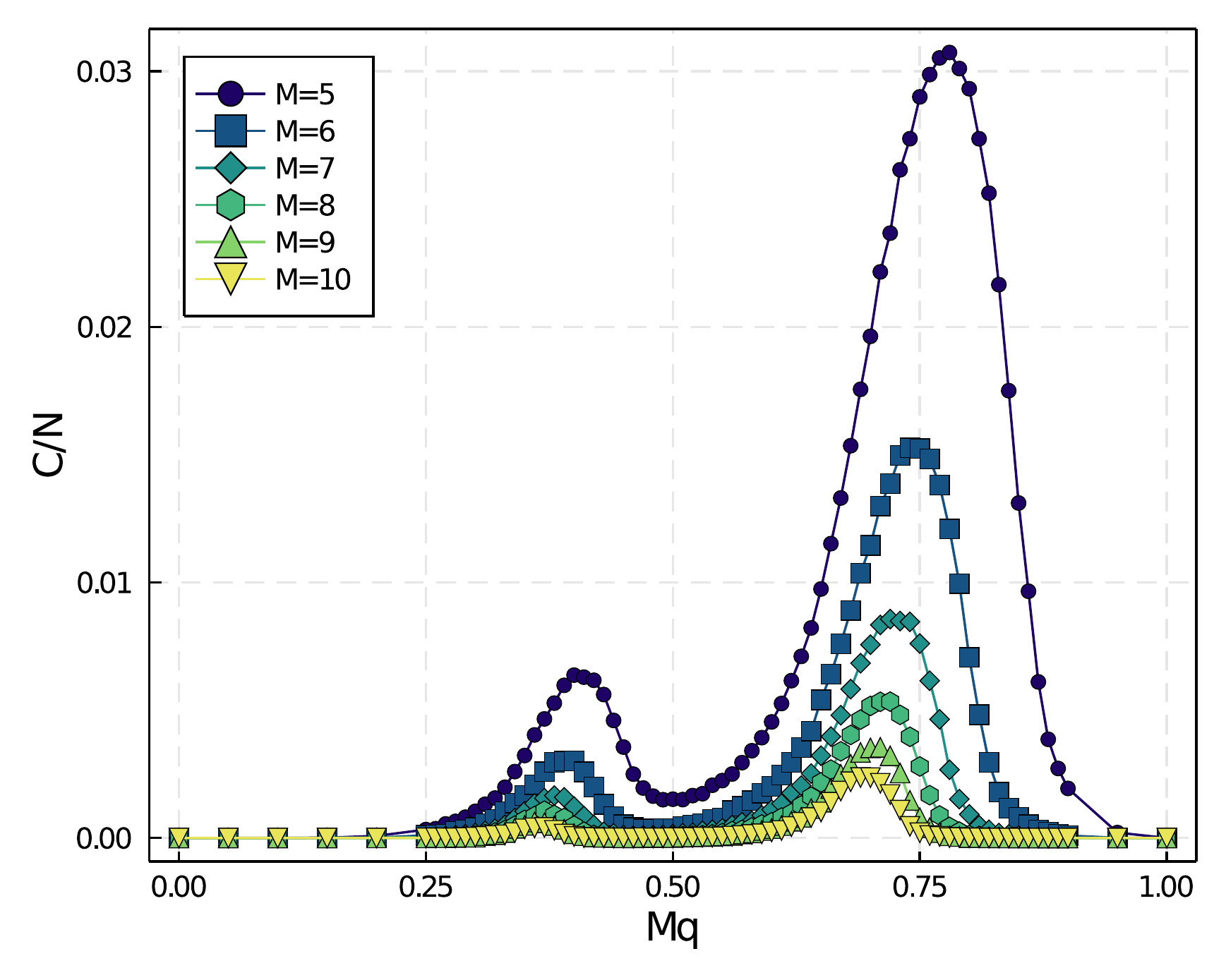}
  \caption{$C(N,M,q)$ as a function of $Mq$ for different $M$s, computed for the the toric code on a hexagonal lattice with  $N=3\times 40\times 40$ physical qubits. The two peaks correspond to the percolation transition of $X$ and $Z$ stabilizers respectively.}
  \label{hex-peaks}
\end{figure}

As discussed in the main text, in the case of toric on code on non self-dual lattices, $X$ and $Z$ stabilizers need to be studied separately as they live of different lattices. For example, on the hexagonal lattice, $Z$ stabilizers form a hexagonal lattice and $X$ stabilizers form a triangular lattice. Following the discussion of Sec. \ref{tpmm}, we expect two different peaks corresponding to bond percolation on the hexagonal and triangular lattice at $p_c=0.653$ and $p_c=0.347$ respectively~\cite{PhysRevA.86.020303,sykes1964exact}. Fig. \ref{hex-peaks} displays numerical results confirming this picture.

\end{document}